\documentclass[aps,twocolumn,nofootinbib,superscriptaddress,preprintnumbers,floatfix]{revtex4}

\usepackage{graphicx,multirow}
\usepackage[usenames,dvipsnames]{color}
\usepackage{amsmath,amsfonts}
\usepackage[hypertex]{hyperref}
\usepackage{slashed}
\usepackage[caption=false]{subfig}

\newcommand{\nn}{\nonumber}
\newcommand{\beqa}{\begin{eqnarray}}
\newcommand{\eeqa}{\end{eqnarray}}

\newcommand{\TeV}{~\text{TeV}}
\newcommand{\GeV}{~\text{GeV}}
\newcommand{\MeV}{~\text{MeV}}
\newcommand{\keV}{~\text{keV}}
\newcommand{\cm}{~\text{cm}}

\newcommand{\Sec}[1]{Sec.~\ref{#1}}

\newcommand{\App}[1]{App.~\ref{#1}}
\newcommand{\Apps}[2]{Apps.~\ref{#1} and \ref{#2}}
\newcommand{\Fig}[1]{Fig.~\ref{#1}}
\newcommand{\Ref}[1]{Ref.~\cite{#1}}
\newcommand{\Refs}[1]{Refs.~\cite{#1}}
\newcommand{\Eq}[1]{Eq.~(\ref{#1})}

\makeatletter
\def\simgt{\mathrel{\lower2.5pt\vbox{\lineskip=0pt\baselineskip=0pt
      \hbox{$>$}\hbox{$\sim$}}}}
\def\simlt{\mathrel{\lower2.5pt\vbox{\lineskip=0pt\baselineskip=0pt
      \hbox{$<$}\hbox{$\sim$}}}}
\makeatother

\catcode`\@=11
\def\overlrarrow#1{\vbox{\ialign{##\crcr
      $\leftrightarrow$\crcr\noalign{\kern-\p@\nointerlineskip}
      $\hfil\displaystyle{#1}\hfil$\crcr}}}
\catcode`@=12

\begin{document}

\title{On dark matter models with uniquely spin-dependent detection possibilities}

\author{Marat Freytsis}
\affiliation{Berkeley Center for Theoretical Physics, Department of Physics,
University of California, Berkeley, CA 94720}
\affiliation{Ernest Orlando Lawrence Berkeley National Laboratory,
University of California, Berkeley, CA 94720}

\author{Zoltan Ligeti}
\affiliation{Ernest Orlando Lawrence Berkeley National Laboratory,
University of California, Berkeley, CA 94720}

\begin{abstract}

With much higher sensitivities due to coherence effects, it is often assumed
that the first evidence for direct dark matter detection will come from
experiments probing spin-independent interactions. We explore models that would
be invisible in such experiments, but detectable via spin-dependent
interactions. The existence of much larger (or even only) spin-dependent
tree-level interactions is not sufficient, due to potential spin-independent
subdominant or loop-induced interactions. We find that, in such a way, most models
with detectable spin-dependent interactions would also generate detectable
spin-independent interactions. Models in which a light pseudoscalar acts as the
mediator seem to uniquely evade this conclusion. We present a particular viable
dark matter model generating such an interaction.

\end{abstract}

\maketitle

\section{Introduction}
\label{sec:intro}

The sensitivity of dark matter (DM) direct detection experiments is undergoing
rapid progress and is expected to continue in the next decade. There are
a number of proposed experiments which will probe complementary aspects of
dark matter properties with much better sensitivities than the existing ones:
DM mass, spin-independent (SI) and spin-dependent (SD) cross sections,
the dependence of the cross sections on the target nuclei, directional
information, etc.

The focus, rightly, is often on the detection of spin-independent DM
interactions, because, due to a coherence effect, the SI interaction cross
section with heavy nuclei is enhanced by $A^2$, the number of nucleons in a
nucleus, and is, therefore, expected in many models to be the dominant interaction
in DM detectors.

There is a good chance that, in the not-too-distant future, direct detection
experiments will be able to extend their sensitivity to cover the full
detectable parameter space for SI cross sections, down to the $10^{-48}\cm^2$
level, below which atmospheric neutrinos constitute an irreducible background.

Prior studies~\cite{Bertone:2007xj,Barger:2008qd,Belanger:2008gy,Cohen:2010gj}
have considered the relationship between SI and SD cross sections, concluding
that the two are typically correlated when a viable dark matter candidate is
present. Most of the discussions have been in the context of the minimal
supersymmetric standard model. (Similar statements have been made about DM
candidates in universal extra dimensions~\cite{Bertone:2007xj} and little Higgs
models~\cite{Belanger:2008gy}, as well.) In general, the common wisdom is that SI
experiments have a much better chance of first direct detection discovery.

The generality of this conclusion cannot be addressed by merely considering
operators; one must explore the underlying models which determine relationships
between operator coefficients. For example, the conclusions stated above
ultimately stemmed from the assumption of DM with electroweak charges, which
generically implies both mediators with at least weak-scale masses to justify
null results thus far and couplings to the Higgs leading to SI signals. Once
this condition is relaxed, the relationship between SI and SD cross sections
becomes weaker, and models in which SD interactions are more easily accessible,
or even the only interaction accessible in direct detection experiments, become
feasible.

Here we point out that in order to impose the last condition, i.e., uniquely SD
detection, the consideration of subleading effects is crucial. Since, due to
coherence effects, SI experiments are more sensitive than SD ones (currently
by 5 orders of magnitude), a loop-induced SI process might be only marginally
more difficult or possibly even as easy to detect than a tree-level SD one.
Upon considering these additional operators, we find that models with light
pseudoscalars are uniquely capable of generically evading such detection
modes.

Although several ingredients of our analysis appear in the
literature~\cite{Agrawal:2010fh,Fan:2010gt}, the impact of light mediators on a
general analysis of operators has not been heretofore discussed, and the effect
of loop corrections on DM scattering has not been considered in this context. In
\Sec{sec:det}, we review current bounds on SI and SD cross sections and the
expected improvements. \Sec{sec:oper} then constitutes the bulk of the paper. We
discuss operators relevant for the detection of DM  particles, including
operators which become important in the case of light mediators. We then
consider which models could generate exclusively SD interactions, and calculate
the loop-induced interactions that would simultaneously be present. In
\Sec{sec:axion} we construct a viable model achieving our goals, in which the SI
interaction is out of reach, but the SD interaction may be detected in future
experiments. \Sec{sec:conc} concludes.

\section{Prospects of Direct Detection}
\label{sec:det}

The best SI bounds come at present from XENON10~\cite{Angle:2007uj},
CDMS~\cite{Ahmed:2009zw}, and XENON100~\cite{Aprile:2010um}, with the highest
sensitivity from XENON100 near $3 \times 10^{-44}\cm^2$ at $50\GeV$. In
general, optimal sensitivity is for DM masses of order the mass of the recoiling
nucleus.  At higher masses, the sensitivity decreases roughly as
$1/m_\text{DM}$. Within the coming years, XENON100, LUX, and SuperCDMS can
improve these bounds down to the $10^{-45}\cm^2$ or possibly near the
$10^{-46}\cm^2$ level. Ultimately multi-ton Xenon or Germanium experiments can
achieve sensitivities to $10^{-47}\cm^2$ or maybe even $10^{-48}\cm^2$, at which
point atmospheric neutrinos form an irreducible
background~\cite{Cabrera:1984rr,Drukier:1986tm,Strigari:2009bq}, and achieving
sensitivity to lower SI DM-nucleon interactions seem unfeasible. 

For SD detection, the best current limit for DM-proton interaction is near $2
\times 10^{-38}\cm^2$ from SIMPLE~\cite{Felizardo:2010mi}, with slightly weaker
bounds from COUPP~\cite{Behnke:2010xt}, KIMS~\cite{Lee.:2007qn}, and
PICASSO~\cite{Archambault:2009sm}, at similar optimal masses as above.  For
DM-neutron cross sections the best bound is from XENON10~\cite{Angle:2008we} at
$5\times 10^{-39}\cm^2$ at optimal sensitivity near {30\GeV}. Within the next
few years, COUPP~\cite{Behnke:20008zza}, PICASSO~\cite{BarnabeHeider:2005pg},
and XENON100 should improve these to $\text{few} \times 10^{-40}\cm^2$, for both
protons and neutrons. These limits could then be extended to near $5 \times
10^{-41}\cm^2$ with experiments such as DMTPC, or to $5 \times 10^{-43}\cm^2$
for a 500\,kg extension of COUPP~\cite{COUPP500}.

Bounds on direct detection cross sections can also come indirectly from other
experiments. One source is from DM annihilation signals from the Sun. The
annihilation at equilibrium is proportional to the rate of DM capture, which is
driven by the same interactions as direct detection. In this case the SI terms
are not so strongly enhanced over the SD ones, since this capture is mostly due
to light nuclei, almost entirely hydrogen and helium. (Some enhancement does
occur due to small amounts of Fe and O, but bounding the SD interaction
neglecting the SI contribution is conservative.)
Super-Kamiokande~\cite{Desai:2004pq} and IceCube~\cite{Abbasi:2009uz} used this
to place limits on SD proton interactions at around $10^{-38}\cm^2$, assuming
annihilations primarily to $b\bar{b}$. Above $m_{\rm DM} \sim 250\GeV$, IceCube
could even place a bound at $2 \times 10^{-40}\cm^2$ if the DM annihilated to
$W^+ W^-$. However, these indirect bounds do not apply in the case of light
mediators which will be discussed below, since if the annihilations proceed
through a light on-shell particle, decays to neither heavy quarks nor $W$ bosons
occur.

Other bounds can be placed from constraints on operators from collider
searches~\cite{Goodman:2010yf,Goodman:2010ku,Bai:2010hh}. In cases where
the mediator can be integrated out, these searches place bounds on interactions
of very light dark matter better than those of direct detection, while remaining
competitive with them for SD interactions of DM that can be directly produced
at the Tevatron. The expected LHC reach is expected to also remain competitive
with direct detection sensitivites of upcoming expierments~\cite{Goodman:2010ku}.
However, for mediators light enough to be produced on-shell, the bound deteriorates
rapidly~\cite{Bai:2010hh}, and is also not applicable for the class of models we
discuss below.

\begin{table}[t!]
\begin{tabular}{cc|cc}
\hline
\multicolumn{2}{c|}{~~Operator~~}  &  ~~SI / SD~  &  ~Suppression~~ \\
\hline
$\mathcal{O}_1^s =$ & $\phi^2\, \bar{q}q $                                              & SI & ---   \\
$\mathcal{O}_2^s =$ & $\phi^2\, \bar{q}\gamma^5 q$                                      & SD & $q^2$ \\
\hline
$\mathcal{O}_3^s =$ & $\phi^\dagger \partial^\mu \phi\, \bar{q} \gamma_\mu q $          & SI & ---   \\
$\mathcal{O}_4^s =$ & $\phi^\dagger \partial^\mu \phi\, \bar{q} \gamma_\mu \gamma^5 q$  & SD & $v^2$ \\
\hline
\end{tabular}
\caption{Operators relevant for scalar dark matter detection.  The suppression
factor given is for the relevant cross section. Operators $\mathcal{O}_3^s$ and
$\mathcal{O}_4^s$ are only allowed for complex scalars.}
\label{tab:soper}
\end{table}

\begin{table}[t!]
\begin{tabular}{cc|cc}
\hline
\multicolumn{2}{c|}{~~Operator~~}  &  ~~SI / SD~  &  ~Suppression~~ \\
\hline
$\mathcal{O}_1^f =$ & $\bar{\chi}\chi\, \bar{q}q$						& SI & ---	\\
$\mathcal{O}_2^f =$ & $\bar{\chi}i\gamma^5\chi\, \bar{q}q$				& SI & $q^2$	\\
$\mathcal{O}_3^f =$ & $\bar{\chi}\chi\, \bar{q}i\gamma^5q$				& SD & $q^2$        \\
$\mathcal{O}_4^f =$ & $\bar{\chi}\gamma^5\chi\, \bar{q}\gamma^5q$				& SD & $q^4$	\\
\hline
$\mathcal{O}_5^f =$ & $\bar{\chi}\gamma^\mu\chi\, \bar{q}\gamma_\mu q$			& SI & ---	\\
\multirow{2}{*}{$\mathcal{O}_6^f =$} & 
\multirow{2}{*}{$\bar{\chi}\gamma^\mu\gamma^5\chi\, \bar{q}\gamma_\mu q$}
		&  SI &  $v^2$	\\
		&  & SD &  $q^2$	\\
$\mathcal{O}_7^f =$ & $\bar{\chi}\gamma^\mu \chi\, \bar{q}\gamma_\mu\gamma^5q $               & SD & $v^2$ or $q^2$ \\
$\mathcal{O}_8^f =$ & $\bar{\chi}\gamma^\mu\gamma^5\chi\, \bar{q}\gamma_\mu\gamma^5q$         & SD & ---          \\
\hline
$\mathcal{O}_9^f =$ & $\bar{\chi}\sigma^{\mu\nu}\chi\, \bar{q}\sigma_{\mu\nu} q $             & SD & ---          \\
$\mathcal{O}_{10}^f =$ & $\bar{\chi}i\sigma^{\mu\nu}\gamma^5\chi\, \bar{q}\sigma_{\mu\nu} q$  & SI & $q^2$        \\
\hline
\end{tabular}
\caption{Operators relevant for fermionic dark matter detection. Operators
$\mathcal{O}_5^f,\ \mathcal{O}_7^f,\ \mathcal{O}_9^f,\ \mathcal{O}_{10}^f$ only
exist if the dark matter is Dirac. Notations as in Table~\ref{tab:soper}.}
\label{tab:foper}
\end{table}

\section{General Considerations}
\label{sec:oper}

\subsection{Operator Analysis}
\label{subsec:oper}

In order to survey possible models, we first identify all operators through
which dark matter may interact with detectors. In doing so, we will see which
interactions give us the signals we are looking for, and which operators need to
be suppressed by small coefficients or forbidden by symmetries. Similar operator
analyses have been considered before in \Refs{Fan:2010gt, Kurylov:2003ra,
Barger:2008qd}. We present it here as a guide to possible types of underlying
structure.

We assume that the mediator is heavy enough so that for the purposes of direct
detection, describing the interaction of dark matter via a contact term is
a reasonable approximation. Beyond this, we want to consider interactions with
dark matter of arbitrary spin, without making additional assumptions, such
as parity conservation. At the structural level of the operators this encompasses
both elastic and inelastic scattering. Having two (or more) fields of different
mass in the DM sector only leads to differences in kinematics and the presence
of operators that are otherwise zero for Majorana fermions and real bosons for
symmetry reasons (discussed below).

The smallest number of operators, as expected, are furnished by scalar dark
matter candidates. These are listed in Table~\ref{tab:soper}. Note that
$\mathcal{O}_3^s$ and $\mathcal{O}_4^s$ are nonvanishing only if the dark matter
candidate is complex.

For fermionic dark matter, the operators are listed in Table~\ref{tab:foper}. If
the dark matter candidate is a Majorana fermion, the operators
$\mathcal{O}_5^f,\ \mathcal{O}_7^f,\ \mathcal{O}_9^f,\ \mathcal{O}_{10}^f$ are
absent, as they are odd under charge conjugation. There are only two operators
with tensor couplings. Since $\sigma^{\mu\nu}\gamma^5 = 
i\epsilon^{\mu\nu\rho\sigma}\sigma_{\rho\sigma}/2$, not all
(pseudo)tensor-(pseudo)tensor combinations are linearly independent. In
addition, $\mathcal{O}_7^f$ has separate SD terms suppressed independently by
$v^2$ and $q^2$, while $\mathcal{O}_6^f$, commonly referred to as the anapole
moment coupling, has contributions to both SI and SD cross sections with
different suppression factors. (Here, as elsewhere in the paper, $v$ is the
velocity of DM in the halo, approximately $10^{-3}$, while $q$ is the momentum
transfer in the interaction.)

Finally, in Table~\ref{tab:voper}, we give the possible operators for vector
dark matter candidates. Similar to the case of scalar dark matter, the operators
$\mathcal{O}_3^v$ and $\mathcal{O}_4^v$ are only present if the vector is
complex.

\begin{table}[tb!]
\begin{tabular}{cc|cc}
\hline
\multicolumn{2}{c|}{~~Operator~~}  &  ~~SI / SD~  &  ~Suppression~~ \\
\hline
$\mathcal{O}_1^v =$ & $B^\mu B_\mu\, \bar{q}q $						& SI & ---       \\
$\mathcal{O}_2^v =$ & $B^\mu B_\mu\, \bar{q}\gamma^5q$					& SD & $q^2$     \\
\hline
$\mathcal{O}_3^v =$ & $B^\dagger_\mu\partial^\nu B^\mu\, \bar{q}\gamma_\nu q$		& SI &  ---      \\
$\mathcal{O}_4^v =$ & $B^\dagger_\mu\partial^\nu B^\mu\, \bar{q}\gamma_\nu\gamma^5 q$	& SD & $v^2$     \\
$\mathcal{O}_5^v =$ & $B^\mu\partial_\mu B^\nu\, \bar{q}\gamma_\nu q$			& SI & $v^2 q^2$ \\
$\mathcal{O}_6^v =$ & $B^\mu\partial_\mu B^\nu\, \bar{q}\gamma_\nu\gamma^5 q$		& SD & $q^2$     \\
\hline
\multirow{2}{*}{$\mathcal{O}_7^v =$} &  \multirow{2}{*}{$\epsilon_{\mu\nu\rho\sigma}B^\mu\partial^\nu B^\rho\, \bar{q}\gamma^\sigma q$}
		& SI  &  $v^2$   \\
		&  & SD  &  $q^2$   \\
$\mathcal{O}_8^v =$ & $\epsilon_{\mu\nu\rho\sigma}B^\mu\partial^\nu B^\rho\, \bar{q}\gamma^\sigma\!\gamma^5 q $  & SD & ---       \\
\hline
\end{tabular}
\caption{Operators relevant for vector dark matter detection. Operators
$\mathcal{O}_3^v$ and $\mathcal{O}_4^v$ only exit for complex vectors fields.
Notations as in Table~\ref{tab:soper}.}
\label{tab:voper}
\end{table}

There are a large number of operators that could mediate SD interactions.
However, for our purposes, some of these may be ignored right away. For
example, $\mathcal{O}_6^f$ and $\mathcal{O}_7^v$ lead to both SD and SI
interactions of comparable magnitudes. It may naively seem that all operators
that come with kinematic suppression factors can be dismissed just as easily.
After all, with DM in the galactic halo at such low velocities,
the nonrelativistic limit is appropriate for detection, and traditionally such
operators have indeed been neglected. Let us examine this assumption more carefully.

Within the dominant WIMP paradigm, the mediator has typically been assumed to be
at the weak scale, with direct detection occurring with ${\cal O}(100\MeV)$
momentum transfers and ${\cal O}(100\keV)$ recoil energies. In that case, the 
integrated-out mediator sets the scale of the operators through a factor of
$1/m_W^2$. In the nonrelativistic limit, terms like $\bar{\psi}\gamma^5\psi$ are
suppressed by factors of $|\vec{q}\,|/2m_N$ or $|\vec{q}\,|/2m_{\rm DM}$, while
others, like $\bar{\psi}\gamma^\mu\gamma^5\psi$, have some components scale as
$v$. Operators with any of these factor can typically be dismissed, because they
are suppressed by $O(10^{3})$. This means that, even if present, such
interactions can be ignored. For example, in the case of Majorana fermion dark
matter, such as the neutralino in supersymmetric models, the only two operators
that need to be considered are scalar\,--\,scalar and
axial-vector\,--\,axial-vector~\cite{Goodman:1984dc, Drukier:1986tm,
Wasserman:1986hh}; all others are highly suppressed.

However, recent interest in explaining various possibly DM-related anomalies
have introduced models with $O(\text{GeV})$ mediator particles. In this case, if
the leading operators were suppressed or forbidden for some symmetry reason, the
traditionally subleading operators could lead to contributions of the correct
magnitude to be accessible to current or future direct detection experiments. As
pointed out in \Ref{Chang:2009yt}, these two statements may in fact be
connected, since the spontaneous breaking of a symmetry forbidding the
appearance of certain operators can provide for a natural explanation for the
presence of light (pseudo) Nambu--Goldstone scalars.

This opens up new possibilities. If SI operators without kinematic suppression
factors are forbidden or highly suppressed for other reasons, the set of
operators which may lead to a detectable SD signal becomes much larger. 

\subsection{Renormalizable Models}
\label{subsec:models}

If we wish to remain agnostic about the nature of the DM-nucleon interactions,
we can say no more. However, if a further step is to be taken, it seems most
conservative to assume that the DM comes from some theory with renormalizable
interactions in which the operators leading to direct detection come from heavy
states that have been integrated out. One can then ask what sort of renormalizable
interactions could lead to the operators given above.
Such a procedure was followed in \Ref{Agrawal:2010fh}. Here we quote
their results, along with the additional possibilities afforded by interactions
yielding kinematically suppressed operators.

For scalar DM, the only option for generating solely SD operators  seems to be a
$t$-channel exchange of a light pseudoscalar, which yields $\mathcal{O}_2^s$.
While such an interaction breaks parity, given that parity is badly broken
already in the standard model (SM), this is not a serious concern.

For fermionic DM, several possibilities present themselves. Once again a
$t$-channel light pseudoscalar exchange produces solely SD interactions via
$\mathcal{O}_4^f$. Additionally, for Majorana fermions, the $t$-channel exchange
of a vector with axial couplings, either the SM $Z$ or a new $Z'$, will generate
only a single kinematically unsuppressed operator, $\mathcal{O}_8^f$. Other
options are an $s$ or $u$-channel coupling through either a scalar or vector,
provided the couplings are chiral, in which case $\mathcal{O}_8^f$ is generated
again. If the couplings are not chiral, $\mathcal{O}_1^f$ is produced as well.

Finally, for vector DM, a light pseudoscalar in the $t$ channel produces only
$\mathcal{O}_2^v$, which breaks parity as in the scalar case.  Alternatively an
$s$ or $u$-channel coupling through a fermion makes $\mathcal{O}_8^v$ the
leading operator, if the coupling is chiral while the vector boson is real.

\subsection{Loops and Subleading Interactions}
\label{subsec:loops}

Suppose that one is presented with a model in which one of the above SD
interactions is the only one present, or dominant over other by many orders of
magnitude. Does that mean that only an experiment sensitive to SD interactions
would see a signal? Not necessarily.

The bounds on SI cross sections are currently 5\,--\,7 orders of magnitude
higher than the SD ones, and this looks to continue to be the case in the
future. Therefore if any of the SD interactions discussed above induce
subleading SI couplings, such an effect could potentially be visible in a SI
experiment. There are two sources for such effects. First, there are
kinematically suppressed contributions of tree level scattering that were
ignored above. These are easily estimated from
Tables~\ref{tab:soper}--\ref{tab:voper} given earlier. Second, the tree-level SD
interactions can induce SI couplings at loop level. These are not as simple to
estimate, and should be calculated to confirm their effect.

\begin{figure}[tb]
  \subfloat[Tree level]{
    \label{fig:Zexchtree}
    \includegraphics[width=0.3\columnwidth]{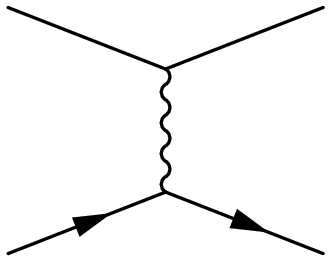}}\hspace*{.5cm}
  \subfloat[Loop processes]{
    \label{fig:Zexchloops}
    \includegraphics[width=0.3\columnwidth]{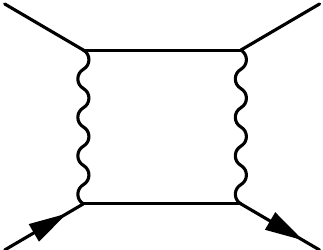}\hspace*{.5cm}
    \includegraphics[width=0.225\columnwidth]{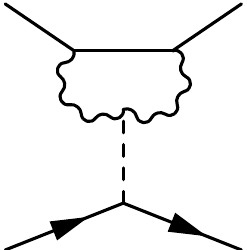}}
  \caption{The tree and loop level contributions to scattering of Majorana
  fermions through a $Z$ boson. For all box diagrams, the crossed box diagram
  is included in calculations but not depicted. In the last diagram, a Higgs
  mediates the scattering through a $Z$ loop.}
  \label{fig:Zexch}
\end{figure}

\begin{figure}[t]
  \subfloat[Tree level]{
    \label{fig:fexchtree}
    \includegraphics[width=0.3\columnwidth]{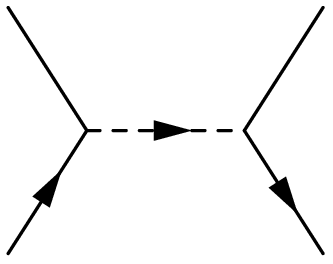}}\hfil
  \subfloat[Loop processes]{
    \label{fig:fexchloops}
    \includegraphics[width=0.3\columnwidth]{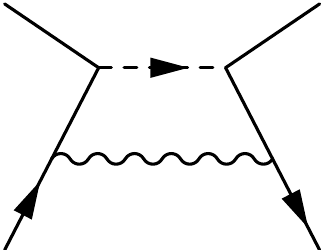}}
  \caption{The tree and loop level contributions to scattering of Majorana
  fermions through a $s$-channel scalar.}
  \label{fig:fexch}
\end{figure}

Let us consider a $Z$ (or $Z'$ exchange) with a Majorana fermion, as in
\Fig{fig:Zexchtree}. While the dominant contribution comes from
$\mathcal{O}_8^f$, also present is $\mathcal{O}_6^f$, the anapole coupling. We
see that this gives rise to a SI interaction suppressed by $v^2$. Similarly, both
the scalar exchange of \Fig{fig:fexchtree} and the equivalent diagram for vector
exchange give an anapole coupling after using Fiertz identities. A fermion
exchange of the same form in the case of vector DM produces $\mathcal{O}_7^v$
as well as $\mathcal{O}_8^v$ in the chiral limit, which again mediates
a $v^2$ suppressed SI coupling. In all of these cases, there is a SI scattering
cross section no more than $O(10^6)$ smaller than the SD one, independent of any
other field content of a model. This means that such interactions would be seen
in SI experiments simultaneously or in the next generation of experiements after
they appear in SD ones. Only the pseudoscalar exchanges evade this, as they lead
to no $v^2$ suppressed subleading contributions to DM-nucleon scattering at all.

All the aforementioned interactions should also be computed at the one-loop
level. While these will be suppressed by loop factors and extra couplings, they
may also generate SI interactions. For large enough couplings, these loops might
even give rise to interactions larger than the kinematically-suppressed ones
discussed above, and so might be even more readily detectable.

Without making any further assumptions about the underlying model, we can
already identify diagrams which will produce SI interactions at loop-level. For
SD interactions involving a $t$-channel exchange, at a minimum, exchanging two
mediators in a box diagram will give rise to a SI interaction. For an $s$ or
$u$-channel processes, a SI loop level contribution can come from a loop with
$W$ or $Z$ bosons exchanged between the quarks.

Consider the exchange of a $Z$ with axial couplings to quarks. (We will discuss
the case of a $Z'$ shortly.) In that case, the quark level operator for tree-level
scattering (\Fig{fig:Zexchtree}) is
\begin{equation}
\frac{g_2^2}{2\cos^2\theta_W}\, T^q_3\, \frac{Q}{2}\, \frac{1}{m_Z^2}\,
  \bar{\chi}\gamma^\mu\gamma^5\chi\, \bar{q}\gamma_\mu\gamma^5q\,,
\end{equation}
where $Q$ is the coupling of the DM to the $Z$. Then the DM-proton SD cross
section generated is (see \Apps{app:matel}{app:xsec} for details)
\begin{equation}
  \sigma_\text{SD}^{\chi p} \approx (1.5 \times 10^{-39}\cm^2)\, 
    \left(\frac{Q}{0.1}\right)^2,
\end{equation}
with the DM-neutron cross sections about 20\% smaller. In this case, two
one-loop processes lead to SI effective interactions: one with two $Z$
exchanges, and a Higgs coupling through a $Z$ loop to the DM
(\Fig{fig:Zexchloops}). We work in the limit $m_q \ll m_Z \ll m_\text{DM}$.
(This limit is generally the one in which the DM has the correct relic abundance
in models where the only coupling of the DM to the quarks is through electroweak
bosons, while foregoing the last inequality only yields $O(1)$ changes, see
\Ref{Essig:2007az}.) The SI contribution to the effective coupling is
then~\cite{Cirelli:2005uq,Essig:2007az}\footnote{In deriving this result, along with those
following, we have set several quark operators, such as
\begin{gather*}
m_q\bar{\chi}\chi\bar{q}q, \quad
\bar{\chi}\chi \bar{q} i\slashed{\partial} q,\\
\frac{4}{3m_\text{DM}}\, \bar{\chi}i\partial_\mu\gamma_\nu\chi\,
  \bar{q}i\left(i\partial^\mu\gamma^\nu + \partial^\nu\gamma^\mu
   - \frac{1}{2}g^{\mu\nu}\slashed{\partial}\right)q,
\end{gather*}
which all simplify to $m_q\, \bar{\chi}\chi\, \bar{q}q$ on shell, but can have
different nuclear matrix elements, to their on-shell value. In fact this seems
to yield a conservative estimate, as out of the nuclear matrix elements known,
the first one has the smallest value (for a detailed discussion of these
issues see \Ref{Drees:1993bu}).}
\begin{equation}
  \frac{1}{4\pi}\, \frac{g^4_2\, Q^2}{\cos^4 \theta_W\, m_Z}
  \left[\frac{(T^q_3)^2}{2m_Z^2}
    +\frac{1}{4m_h^2}\right] m_q\, \bar{\chi}\chi\, \bar{q}q\,.
\end{equation}
Taking a reference value of $m_h = 120\GeV$, this interactions will induce
a SI cross section of
\begin{equation}
  \sigma_\text{SI}^{\chi N} = (4 \times 10^{-47}\cm^2)\,
    \left(\frac{Q^2}{0.1}\right)^2 .
\end{equation}
Asking that the SD signal be just beyond current SD experimental bounds implies
$Q \sim 0.3$, giving a SI cross section of $4 \times 10^{-47}\cm^2$. This, while
not detectable in experiments underway, is feasible with ones in preparation.

This result would make the $v^2$-suppressed constribution to SI scattering
dominant. However, it is worth mentioning that this cross section acts as
a lower bound --- it could be that the DM particle is part of larger
representation of $SU(2)$, in which case additional loops involving $W$s would
also contribute. Generally, the size off the cross section grows as $n^2$,
with $n$ the dimension of the representation~\cite{Cirelli:2005uq}, making it
possible for the loop contribution to be dominant, and not merely competitive
with the kinetically suppressed contribution, and even being large enough to be
discovered simultaneously with the SD signal.

If one wishes to consider models with a new $Z'$, then the existence of a $Z'$
with Higgs coupling becomes model dependent. To talk about a lower bound, we can
then ignore the contribution of the second term in the effective coupling. The
heavier mediator mass that such a model would entail would have to be offset
with a larger coupling in order to be detectable. Thus, at loop level, one would
generally expect the effective interaction to be of at least similar size, or
possibly larger, due to the higher power of the coupling appearing in the
loop-induced term.

If one considers the possibility of a light $Z'$, which are not ruled out by
collider constraints down to the GeV range for gauge couplings smaller than the
SM by $10^{-2}$, the situation discussed above would be reversed, and one would
expect a smaller loop-induced contribution. However, the SI contribution due to
kinematically suppressed operators is insensitive to changes in the mediator
mass, and would still be present. Constructing a model without such operators
and without significant fine-tuning seems extremely difficult. It is difficult
to say more in generality, due to the large freedom in assigning masses and
charges under a new gauge group.

\begin{figure}[t!]
  \subfloat[Tree level]{
    \label{fig:pseudtree}
    \includegraphics[width=0.3\columnwidth]{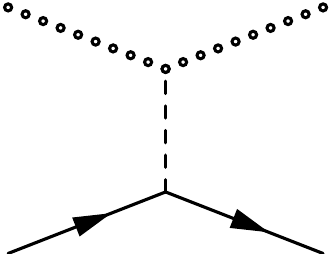}}\hfil
  \subfloat[Loop processes]{
    \label{fig:pseudloops}
    \includegraphics[width=0.3\columnwidth]{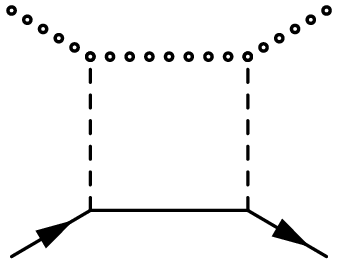}}
  \caption{The tree and loop level contributions to scattering DM mediated by
  a light pseudoscalar. The dotted line can represent either a scalar, fermion,
  or vector boson.}
  \label{fig:pseudexch}
\end{figure}

Now let us consider DM with chiral couplings to the SM via an $s$ or $u$-channel.
The most model-independent loop-level processes here come from box diagrams
with the quarks exchanging a $W$ or $Z$ boson, an example of which is given
in \Fig{fig:fexchloops}. The contributions of the loops have completely different
forms depending on whether the coupling of the DM is left- or right-handed.
However, in all cases the loop-level processes only give rise to suppressed
SD contributions. In addition to DM of the form in \Fig{fig:fexch}, this is also 
true for the cases of fermionic DM with a vector mediator and vector DM with
a fermion mediator of similar topologies. In this case, we find that the most
reliable lower bound on a SI cross section in this case comes from the $v^2$
suppressed contribution to the tree-level interaction discussed earlier.

Finally, let us turn to the box diagrams induced in the cases of light
pseudoscalar exchange, \Fig{fig:pseudloops}. First, we consider the case of scalar
DM. At tree level, the operator obtained after integrating out the pseudoscalar is
\begin{equation}
  \frac{1}{m_a^2}\, \xi\, y_q\, m_\phi\, \phi^\dagger \phi\, \bar{q}i\gamma^5 q,
\end{equation}
where $y_q$ is the Yukawa coupling of the quark, so $\xi$ absorbs both the
coupling of the DM and mediator, and any scaling to Yukawas of the mediator-quark
coupling. This leads to a a tree-level cross section of
\begin{equation}
  \sigma_\text{SD}^{\phi p} \approx (8 \times 10^{-37}\cm^2)
    \left(\frac{\xi}{0.1}\right)^2 \left(\frac{1\GeV}{m_a}\right)^4.
\label{eq:scxsec}
\end{equation}
(See \App{app:xsec} for the definition of the cross section in cases of
kinematically suppressed operators.) For a mediator with mass of a few GeV
and $\xi = 0.01$, this would be accessible to currently running searches.

The calculation of the loop diagram in the same limits as the previous
$Z$-mediated case does not give as compact of an answer, but can be expressed in
closed form in terms of Passarino--Veltman scalar
integrals~\cite{Passarino:1978jh}, computed with the use of
FeynCalc~\cite{Mertig:1990an} as
\begin{widetext}
\begin{equation}
\begin{split}
\frac{1}{(4\pi)^2}\, \xi^2\, y_q^2\, & \Big[
  C_0(m_\phi^2,0,m_\phi^2;m_\phi^2,m_a^2,0)
  - C_0(m_\phi^2,m_\phi^2,0;m_a^2,m_\phi^2,m_a^2)\\
& + m_a^2\, D_0(m_\phi^2,m_\phi^2,0,0,0,m_\phi^2;m_a^2,m_\phi^2,m_a^2,0)\Big]\,
  \phi^\dagger \partial^\mu \phi\, \bar{q}\gamma_\mu q.
\end{split}
\end{equation}
A numerical evaluation of the coefficients show the $C_0$ and $D_0$ functions
with these parameters to scale as $\ln(m_a/m_\phi)$ and $\ln^2(m_a/m_\phi)$,
respectively, beyond their overall $1/m_\phi^2$ dependence. Using a fiducial
value of $m_a/m_\phi = 0.01$ gives
\begin{equation}
  \frac{1}{(4\pi)^2}\, \frac{\xi^2\, y_q^2}{m_\phi^2}\, C_S\,
    \phi^\dagger \partial^\mu \phi\, \bar{q}\gamma_\mu q\,,
\end{equation}
where $C_S \approx 80$. Note that if the DM were real, this operator vanishes
identically, and there is no loop induced coupling at one loop order at all. If
present, the cross section induced is
\begin{equation}
  \sigma_\text{SI}^{\phi N} \approx (4 \times 10^{-54}\cm^2)
    \left(\frac{\xi}{0.1}\right)^4 \left(\frac{100\GeV}{m_\phi}\right)^4,
\label{eq:scloopxsec}
\end{equation}
undetectable for any any choice of parameters that would make the SD cross
section detectable.

The case of vector DM is very similar. For
\begin{equation}
  \frac{1}{m_a^2}\, \xi\, y_q\, m_B\, B_\mu^\dagger B^\mu\, 
  \bar{q}i\gamma^5 q,
\end{equation}
the tree-level cross section takes the same value as \Eq{eq:scxsec}. Meanwhile,
the loop induced coupling is
\begin{equation}
\begin{split}
&\frac{1}{(4\pi)^2}\, \xi^2\, y_q^2\, \bigg\{ C_0(m_B^2,0,m_B^2;m_B^2,m_a^2,0)
  - C_0(m_B^2,m_B^2,0;m_a^2,m_B^2,m_a^2)\\
&\quad + m_a^2\, D_0(m_B^2,m_B^2,0,0,0,m_B^2;m_a^2,m_B^2,m_a^2,0)
  + \frac{1}{4 m_B^2}\Big[B_0(m_B^2;m_a^2,m_B^2) - B_0(m_B^2;0,m_B^2)\Big]
  \bigg\}\,
  B_\nu^\dagger\, \partial^\mu B^\nu\, \bar{q}\gamma_\mu q\,,
\end{split}
\end{equation}
which numerically evaluates to
\begin{equation}
  \frac{1}{(4\pi)^2} \frac{\xi^2\, y_q^2}{m_B^2}\, C_V
    B_\nu^\dagger\, \partial^\mu B^\nu\, \bar{q}\gamma_\mu q\,,
\end{equation}
with $C_V \approx 80$ very close to the scalar case, giving a loop induced
SI cross section as in \Eq{eq:scloopxsec}, and similarly giving no contribution
if the DM were real.

The case of fermionic DM is slightly different. This is because the tree-level
operator responsible for scattering is
\begin{equation}
  \frac{1}{m_a^2}\, \xi\, y_q\, \bar{\chi}i\gamma^5\chi\, \bar{q}i\gamma^5 q,
\end{equation}
and, therefore, is parametrically suppressed by $q^4$, instead of the
previous cases' $q^2$. The tree level cross section then becomes
\begin{equation}
  \sigma_\text{SD}^{\chi p} \approx (3 \times 10^{-43}\cm^2)
    \left(\frac{\xi}{0.1}\right)^2 \left(\frac{1\GeV}{m_a}\right)^4.
\end{equation}
We see that due to the greater momentum suppression, we require a lighter
mediator mass and cannot afford the coupling of the DM to be as small as in the
bosonic case above. In this case a cross section detectable in current
experiments would require, for example, a mediator with $m_a = 100\MeV$ and
$\xi= 0.1$.

Meanwhile, the effective coupling from computing the loop diagram in the same
limits as the other cases is
\begin{equation}
\begin{split}
  \frac1{(4\pi)^2} & \frac{\xi^2\, y_q^2}{m^2_\chi} \bigg\{ \bigg[
    \frac12 + \frac{m_\chi^2}2\, C_0(m_\chi^2,m_\chi^2,0;m_a^2,m_\chi^2,m_a^2)
    - m_\chi^2\, C_0(0,m_\chi^2,m_\chi^2;0,m_a^2,m_\chi^2) \bigg]\,
    \bar{\chi}\gamma^\mu\chi\, \bar{q}\gamma_\mu q\\
  & + \frac{3}{8} \Big[1 + B_0(m_\chi^2;0,m_\chi^2) - B_0(0;m_a^2,m_a^2) 
    + 4 m_\chi^2\, C_0(m_\chi^2,0,m_\chi^2;m_\chi^2,m_a^2,0)
    - m_\chi^2\, C_0(m_\chi^2,m_\chi^2,0;m_a^2,m_\chi^2,m_a^2)\\
  & + 3 m_a^2 m_\chi^2\,
  D_0(m_\chi^2,m_\chi^2,0,0,0,m_\chi^2;m_a^2,m_\chi^2,m_a^2,0) \Big]\,
    \frac{m_q}{m_\chi}\, \bar{\chi}\chi\, \bar{q} q \bigg\}\,
\end{split}
\end{equation}
\end{widetext}
which numerically yields
\begin{equation}
  \frac{1}{(4\pi)^2}\, \frac{\xi^2\, y_q^2}{m^2_\chi}
    \left( C_{F_1}\, \bar{\chi}\gamma^\mu\chi\, \bar{q}\gamma_\mu q
    + C_{F_2}\, \frac{m_q}{m_\chi}\, \bar{\chi}\chi\, \bar{q} q\right),
\end{equation}
with $C_{F_1} \approx 4.8$ and $C_{F_2} \approx 170$. The magnitudes of these
coefficients can be understood as arising from the  $\ln(m_a/m_\phi)$ and
$\ln^2(m_a/m_\phi)$ behavior of $C_0$ and $D_0$ mentioned above. The
loop-level cross section is then
\begin{equation}
  \sigma_\text{SI}^{\chi N} \approx (3 \times 10^{-56}\cm^2)
  \left(\frac{\xi}{0.1}\right)^4 \left(\frac{100\GeV}{m_\chi}\right)^4.
\end{equation}
We will confirm below in the explicit model of \Sec{sec:axion} that the loop
induced coupling is indeed tiny, but it is simple to see here why this is
generically so.

Unlike in the massive mediator cases, there are two mass scales in the dark
sector, that of the DM itself and that of the mediator. At tree level, the
lighter mediator mass is the one that appears in the denominator of the
operator. However, at loop level, the value of the loop integral is
parametrically controlled by the mass of the DM, the heaviest particle in the
loop. Additionally, a pseudoscalar which is the Nambu--Goldstone boson of a
broken symmetry would be expected to couple to quarks proportional to the masses
of the quarks. Thus, at loop level, the effective operator would be expected to
be suppressed by extra factors of quark Yukawa couplings. Together, both effects
combine to make the loop-level coupling to be as many as 20 orders smaller than
the tree level one, with higher order corrections to the non-relativistic scattering
approximation coming at similar orders as $q^2v^4$, so that the SI induced interaction
is expected to be completely negligible.

\section{The Axion Portal}
\label{sec:axion}

We have just seen that without tuning of couplings, models with light
pseudoscalar mediators provide the unique method of avoiding any SI signal,
while still producing a SD direct detection signature. Now we turn to the
question of whether a viable model producing DM with the correct abundance can
have these features.

Coupling a light pseudoscalar to quarks is most efficiently achieved by adding a
scalar field which spontaneously breaks a global symmetry, and which, by mixing
with the Higgs, gets a coupling to the SM. Allowing this scalar to have a new
global charge, while adding new fermions charged under the same symmetry,
ensures that the new scalar field is the only method for the new fermions to
interact with the SM.

As a simple realization of such a mechanism, where the dominant interaction
is $\mathcal{O}^f_4$ via a pseudoscalar interaction, we introduce, following
\Ref{Nomura:2008ru}, a scalar field charged under a new global $U(1)_X$ charge
that is spontaneously broken to
\begin{equation}
  S = \left(f_a + \frac{s}{\sqrt{2}}\right)\,
  \exp\bigg(\frac{ia}{\sqrt{2}f_a}\bigg) \,.
\end{equation}
This scalar field is coupled to a new fermion, which is vector-like under
the SM, through $\mathcal{L} = -\xi S\chi\chi^c + \text{h.c.}$, so that after
the scalar field acquires a vacuum expectation value, the fermion receives
a mass of $m_\chi = \xi f_a$, allowing it to act as dark matter,
with stability ensured by the remnant of $U(1)_X$ after breaking.

In order for the pseudoscalar to interact with the SM, some known particles
must also carry charges under the new $U(1)_X$. In a two Higgs doublet model,
this can be accomplished by adding a term of the form
\begin{equation}
  \mathcal{L} = \lambda S^n H_u H_d + \text{h.c.},
\end{equation}
by assigning the appropriate charges to the Higgses and SM fermions, and
promoting the $U(1)_X$ to a Peccei-Quinn (PQ) symmetry. For $n=2$, this
coupling is of the same form as in the case of the DFSZ
axion~\cite{Dine:1981rt,Zhitnitsky:1980tq}, while the $n=1$ case functions like
that of the PQ-symmetric limit of the next-to-minimal supersymmetric standard
model~\cite{Hall:2004qd} We now have a dark matter candidate coupling to the SM
though a massive scalar and an axionlike Nambu--Goldstone boson. The
Nambu--Goldstone boson is assumed to get a small mass through an unspecified
mechanism. Anticipating making the scalar heavy, by virtue of
\begin{equation}
  \langle\sigma v\rangle_{\chi\chi^c \to sa} = \frac{m_\chi^2}{64\pi f_a^4}
    \left(1 - \frac{m_s^2}{4m_\chi^2}\right)+ O(v^4) \,,
\end{equation}
a choice of, say, $m_s = f_a = 1\TeV$ and  $m_\chi = 1.1\TeV$ (corresponding to
$\xi = 1.1$)  yields a cross section of $3 \times 10^{-26} \cm^3/ \text{s}$ and
so generates the correct order of magnitude for the relic
abundance~\cite{Nomura:2008ru}.

For direct detection, two channels present themselves. The scalar gives a SI
cross section through the operator $\mathcal{O}^f_1$, due to mixing of the
scalar with the two $CP$-even Higgses, while the light axionlike state yields a
SD interaction, $\mathcal{O}^f_4$, by a similar mixing with the $CP$-odd Higgs.
For our purposes, we need check whether this tree-level SI cross section can be
small enough to be completely negligible.

The mixing of the scalar with the two $CP$-even Higgs has a lot of arbitrariness
to it due to the 11 constants in the most general $U(1)_\text{PQ}$-preserving
two-Higgs-doublet and one-singlet potential. However, we can say that barring
accidental cancellations, this mixing will be $\epsilon = O(v_\text{ew}/f_a)$,
so that we may write the tree-level SI cross section as
\begin{equation}
  \sigma_\text{SI}^{\chi N} \approx (2 \times 10^{-42}\cm^2)\, 
  \xi^2\, \epsilon^2 \left(\frac{100\GeV}{m_s}\right)^4 ,
\end{equation}
(See \App{app:matel} for a caveat on the values of the nuclear matrix elements
in this calculation.) In the model considered in \Ref{Nomura:2008ru}, $m_s$
needed to be light, $O(10\GeV)$, in order to provide a mechanism for Sommerfeld
enhancement to explain astrophysical anomalies. In that case, the direct detection
cross section was in tension with the SI bound and could only be slightly beyond
current limits, at $\text{a few}\times 10^{-43}\cm^2$. However, if we impose no
such condition, $m_s$ could be larger. If it is at the electroweak scale, then
the cross section is at most $\text{a few} \times 10^{-45}\cm^2$, smaller than
the sensitivity of the next generation of direct detection experiments. If $m_s
\sim O(1\TeV)$, a reasonable choice given the scale of $f_a$ in this setup,
then the cross section becomes undetectably small, below the irreducible
atmospheric neutrino limit.

Let us next consider the pseudoscalar channel. With the interaction kinetically
suppressed by the momentum transfer as $q^4$, we cannot merely compute the
cross section in the limit of $q^2 \to 0$ as we did in the scalar exchange
case. Instead, we must define a cross section at a fixed momentum transfer.
(See \App{app:xsec} for a more thorough discussion.) We choose to do so at
$q_\text{ref}^2 = (100 \MeV)^2$. Because the signal is different from that of
unsuppressed interactions relative to the expected recoil energies, the
sensitivities of experiments are modified. This was studied in
\Ref{Chang:2009yt}, with the result that at the same reference momentum
transfer, optimal sensitivities of SD experiments to pseudoscalars remained at
the same order of magnitude as  in the unsuppressed case, but with
$1/m_\text{DM}$ scaling of the limits. 

With this definition, we can compute the SD cross section for $q^2 =
q_\text{ref}^2$ as
\begin{equation}
  \sigma_\text{SD}^{\chi p} \approx (2 \times 10^{-37} \cm^2)\,
  \xi^2 \sin^2\theta\, 
  \frac{q_\text{ref}^2}{4m_\chi^2}
  \bigg(\frac{1\GeV}{m_a}\bigg)^4,
\end{equation}
where $\tan\theta = n \sin 2\beta\, [v_\text{ew} / (2f_a)]$ is the mixing of
the $s$ with the Higgses~\cite{Freytsis:2009ct}. From this we see that given
a DM mass $m_\chi = 1.1\TeV$, a pseudoscalar with a mass $m_a \approx 300\MeV$
generates a cross section of $3 \times 10^{-40}\cm^2$, within the range of
the next generation of direct SD detection experiments. In fact, in a two Higgs
doublet model like this, the nuclear matrix element also has a dependence on
$\beta$ as up-type quarks couple with a coefficient proportional to $\cot\beta$,
while down type ones couple proportional to $\tan\beta$. We have evaluated the
matrix elements for the above cross section at $\tan\beta = 1$. At large values
of $\tan\beta$ the cross section can rise by almost 2 orders of magnitude.

Given the tiny size of the tree-level SI cross section, and in keeping with the
discussion of the previous section, we should confirm that the loop-induced
couplings fail to produce a detectable SI cross section. The calculation mostly
mirrors that of \Sec{subsec:loops}. The only substantial difference is
the aforementioned different coupling to up and down type quarks. As before,
we evaluate the nuclear matrix elements at $\tan\beta = 1$, but this time, varying
$\tan\beta$ cannot only modify the cross section by a factor of $O(1)$ 
as the suppression of $\sin^2\theta$ at high $\tan\beta$ is too strong, so we find
\begin{equation}
  \sigma_\text{SI}^{\chi N} \approx (3 \times 10^{-56}\cm^2)
  \left(\frac{\xi\,\sin\theta}{0.1}\right)^4 \left(\frac{100\GeV}{m_\chi}\right)^4
\end{equation}
with no additional implicit $\tan\beta$ dependence.

\section{Conclusions}
\label{sec:conc}

As the sensitivity of both SI and SD direct DM detection experiments increases,
it is worth asking to what extent the discovery potential of the two methods is
complementary. In this work we have pointed out that when one considers the full
range of possible mediators, instead of being confined to new weak-scale
particles, the range of possible viable interactions generating SD cross
sections increases. At the same time, when one searches for interactions for
which SD experiments are complimentary for discovery --- ones which could not be
seen in any SI experiments without the need for accidental cancellations or
other tuning ---  it becomes necessary to take into account subleading
contributions to scattering, such as suppressed operators and loop processes.
The outcome is that the traditional models considered also generically produce
SI interactions whose suppression is counterbalanced by the greater sensitivity
of SI experiments. The list of viable candidates whose interaction with the SM
can be described by tree-level mediators integrated out in a renormalizable
model is then reduced to merely ones mediated by light pseudoscalars.

We have presented a realistic model of such interactions that generates the right
DM abundance with a fermionic DM candidate without having other interactions
generating detectable SI interactions. 

Similar scenarios can also be considered with a scalar or vector dark matter
candidate. Just as in the case of fermionic DM, $\mathcal{O}_1^{s,v}$ gives the
leading interaction in the nonrelativistic limit, while $\mathcal{O}_2^{s,v}$
is kinematically suppressed. The necessary couplings between the pseudoscalar and
the scalars or vectors cannot be generated in as simple a manner as those used
above, so more model building will be required. However, the suppression is only
by $q^2$, so the mass differences between the scalar and pseudoscalar do not
have to be quite as large, and the couplings themselves can be smaller, so that
the parameter space of couplings and the pseudoscalar mass are not as tightly
limited by experiment, potentially making the exercise worthwhile.

\begin{acknowledgments}

We thank Jeremy Mardon and  Michele Papucci for helpful discussions, and Tomer
Volansky for drawing our attention to the importance of loop contributions.
This work was supported in part by the U.S.\ Department of
Energy under contract DE-AC02-05CH11231.

\end{acknowledgments}

\appendix

\section{Nuclear Matrix Elements}
\label{app:matel}

Here we summarize how to compute the dark matter-nucleon interaction cross
sections from quark-level interactions. Much of this has been discussed in the
DM literature, with the exception of the pseudoscalar matrix element, as it only
plays a role in momentum suppressed cross sections.

For a vector coupling, nuclear matrix elements are straightforward to compute,
since a vector coupling to quarks is a conserved current, so the coupling to a
nucleon is obtained from the sum of the currents of the valence quarks.

In the case of a scalar coupling to quarks, we are interested in the effective
nucleon coupling induced by a quark level coupling:
\begin{equation}
a_q\, m_q\, \bar{q}q \quad \to \quad f_N\, m_N\, \bar{N}N.
\end{equation}
We define the nuclear matrix elements conventionally by
\begin{equation}
  \langle N | m_q \bar{q}q | N \rangle = m_N f^{(N)}_{Tq}.
\end{equation}
On including the coupling to gluons induced by integrating out heavy
quark loops, $f_N$ is given by
\begin{equation}
  f_N = \sum_{q = u,d,s}\! f^{(N)}_{Tq}\, a_q
  + \frac{2}{27}\, f^{(N)}_{TG} \sum_{q = c,b,t} a_q \,,
\end{equation}
where $f^{(N)}_{TG} = 1 -  \sum_{q=u,d,s} f^{(N)}_{Tq}$.

Unlike the $u$ and $d$ matrix elements, which can be extracted from $\pi N$
scattering, the uncertainty associated with the strange quark matrix element
$f^{(N)}_{Ts}$ is higher, which introduces a substantial uncertainty in the
SI coupling to nucleons.  Most studies use numerical values $f^{(N)}_{Ts} \gg
f^{(N)}_{Tu,d}$ based on older calculations. A representative set of values is
that used by the DarkSUSY package~\cite{Gondolo:2004sc}, wherein,
\begin{align}
f^{(p)}_{Tu} &=0.023\,, \quad f^{(p)}_{Td} =0.034\,, \quad f^{(p)}_{Ts} =0.14\,,
\nn\\
f^{(n)}_{Tu} &=0.019\,, \quad f^{(n)}_{Td} =0.041\,, \quad f^{(n)}_{Ts} =0.14\,.
\end{align}
These are the values used for the numerical estimates given above, and in most
of the literature. However, recent lattice QCD results give substantially
smaller values, $f^{(N)}_{Ts} = 0.013 \pm 0.020$~\cite{Takeda:2010id} (see
also~\cite{Young:2009zb,Toussaint:2009pz}), and so the SI cross section from
scalar exchange (if it couples proportionally to mass) may be smaller by a
factor of $2$\,--\,$5$ than numerical results quoted by many calculations.

For SD interaction we need to consider the nuclear matrix elements induced by
the quark level axial-vector and pseudoscalar couplings,
\begin{equation}\label{SDaxialvector}
d_q\, \bar q \gamma_\mu \gamma^5 q \quad \to \quad a_N\, \bar N s^{(N)}_\mu N,
\end{equation}
and
\begin{equation}\label{SDpseudoscalar}
c_q\, m_q\, \bar q i\gamma^5 q \quad \to \quad g_N\, m_N\, \bar N i\gamma^5 N.
\end{equation}
For the axial-vector current, defining
\begin{equation}
  \langle N | \bar{q} \gamma_\mu \gamma^5  q | N \rangle 
  = s^{(N)}_\mu \Delta q^{(N)},
\end{equation}
where $s^{(N)}_\mu$ is the spin of the nucleon, we have
\begin{equation}
  a_N = \sum_{q = u,d,s}\! d_q\, \Delta q^{(N)}.
\end{equation}
The matrix elements coming from polarized  deep inelastic scattering carry much
smaller uncertainties than for the scalar SI interaction above.  For our
numerical results, we use again the DarkSUSY values,
\begin{align}
\Delta u^{(p)} &= \Delta d^{(n)} = 0.77\,, \nn\\
\Delta d^{(p)} &= \Delta u^{(n)} = -0.40\,, \nn\\
\Delta s^{(p)} &= \Delta s^{(n)} = -0.12\,.
\end{align}
More recent determinations favor slightly different values, and the PDG quotes
$\Delta s^{(n)} = -0.09$, $\Delta d^{(n)} = 0.84$, $\Delta u^{(n)} = -0.43$,
with a 0.02 uncertainty for each~\cite{Nakamura:2010zzi}; the effect on our
numerical results is negligible.

For the pseudoscalar current in Eq.~(\ref{SDpseudoscalar}) the nucleon-level
coupling is determined by the same axial-vector matrix elements above. The
relationship is established through generalized Goldberger--Treiman relations.
While not normally considered in dark matter detection, it has been well-studied
in the axion literature~\cite{Srednicki:1985xd,Donnelly:1978ty}.
Taking divergences of the axial currents and using the equations of motion for
the quarks yields~\cite{Cheng:1988im}
\begin{align}
g_N &= (c_u - \bar{c}_q \eta)\, \Delta u^{(N)} \nn\\
  & \quad + (c_d - \bar{c}_q \eta z)\, \Delta d^{(N)}
    + (c_s - \bar{c}_q\eta w)\, \Delta s^{(N)} \,,
\end{align}
where $\eta = (1 + z + w)^{-1}$, $z = m_u/m_d$, and $w = m_u/m_s$, while $\bar{c}_q$
is the mean of the quark coupling coefficients. Due to uncertainties in the value
of $z$, the value of $g_N$ can vary by as much as a factor of $2$.

\section{Cross Sections}
\label{app:xsec}

In this Appendix, we provide a summary of cross sections for DM-nucleon
interactions relevant for calculating the various cross sections discussed above
in the non-relativistic limit.

We first consider the unsuppressed operors in the limit of zero momentum
transfer.  SI cross sections can come from either scalar or vector quark
couplings. Effective DM-nucleon scalar interactions for fermions of the form
\begin{equation}
  f_N\, \bar{\chi}\chi\, \bar{N}{N},
\end{equation}
which are derived from the quark-level couplings using nuclear matrix elements, 
as explained in \App{app:matel}, lead to a DM-nucleus cross section
\begin{equation}
  \hat\sigma = \frac{4}{\pi}\, \hat\mu^2 \big[Z f_p + (A-Z) f_n\big]^2 ,
\end{equation}
for Majorana DM fermions. (For Dirac fermions, all results for Majorana fermions
are divided by 4.) Here $\hat\mu$ is the reduced mass of the DM-nucleus system.
The per-nucleon cross section, which is usually quoted for comparisons, is
\begin{equation}
  \sigma = \frac{4}{\pi}\, \mu^2 \frac{1}{A^2} \big[Z f_p + (A-Z) f_n\big]^2,
\end{equation}
where $\mu$ is the reduced mass of the DM-nucleon system.

For scalar or vector dark matter, the relevant operators are (we include the DM
mass to give all operators the same dimension)
\begin{equation}
  f_N\, m_\phi\, \phi\phi\, \bar{N}N \quad \text{or} \quad 
  f_N\, m_B\, B^\mu B_\mu\, \bar{N} N,
\end{equation}
the nucleon cross section for either operator is
\begin{equation}
  \sigma = \frac{1}{\pi}\, \mu^2 \frac{1}{A^2} \big[Z f_p + (A-Z) f_n\big]^2.
\end{equation}

Vector interactions for fermions only exist in the case of Dirac DM:
\begin{equation}
  b_N\, \bar{\chi}\gamma^\mu\chi\, \bar{N}\gamma_\mu N,
\end{equation}
where, $b_p = 2b_u + b_d$ and $b_n = b_u + 2b_d$, due to vector current
conservation, as discussed above in \App{app:matel}.  Then
\begin{equation}
  \sigma = \frac{1}{\pi}\, \mu^2 \frac{1}{A^2} \big[Z b_p + (A-Z) b_n\big]^2.
\end{equation}
For the operators
\begin{equation}
  b_N\, \phi^\dagger\partial_\mu\phi\, \bar{N}\gamma^\mu N \quad \text{or} \quad
  b_N\, B_\nu^\dagger \partial_\mu B^\nu\, \bar{N}\gamma^\mu N ,
\end{equation}
which only exist for complex scalars or vectors, the cross section is
\begin{equation}
  \sigma = \frac{1}{\pi}\, \mu^2 \frac{1}{A^2} \big[Z b_p + (A-Z) b_n\big]^2 .
\end{equation}

Unsuppressed SD interactions come solely from the quarks' axial currents.
In the case of
\begin{equation}
  a_N\, \bar{\chi}\gamma^\mu \gamma^5 \chi\, \bar{N} \gamma_\mu \gamma^5 N,
\end{equation}
the DM-nucleus cross section is
\begin{equation}
  \hat\sigma = \frac{16}{\pi}\, \hat\mu^2\, a_N^2\, J_N (J_N + 1),
\end{equation}
and for a nucleon
\begin{equation}
  \sigma = \frac{12}{\pi}\, \mu^2\, a_N^2 .
\end{equation}
The only other unsuppressed SD interaction is for vector DM and comes from
\begin{equation}
  a_N\, \epsilon^{\mu\nu\sigma\rho} B_\mu \partial_\nu B_\sigma\,
  \bar{N} \gamma_\rho \gamma^5 N\,.
\end{equation}
Here, the DM-nucleon cross section is
\begin{equation}
  \sigma = \frac{2}{\pi}\, \mu^2\, a_N^2 .
\end{equation}

All of the above cross sections are quoted in the $q^2 \to 0$ limit. In this
limit, interactions mediated by light pseudoscalars are all zero, so we need
another way of expressing such cross sections. To do so, we will use the fact
that while in the nonrelativistic limit $\bar{\psi}\psi \sim 2m$,
$\bar{\psi}\gamma^5\psi \sim q^i \xi^\dagger \sigma^i \xi$, so that using the
results above we can write (since $q^2 \approx |\vec{q}\,|^2$ in the
nonrelativistic limit).
\begin{equation}
 \bar{\chi}\gamma^5\chi\, \bar{N}\gamma^5 N \sim
 \frac{q^2}{4m_\chi^2}\, \frac{q^2}{4m_N^2}\,
 \bar{\chi}\gamma^\mu \gamma^5 \chi\, \bar{N} \gamma_\mu \gamma^5 N\,.
\end{equation}
We then compute the cross section as above, and quote a result at a reference
value of $q^2$. We have chosen $q^2 = (100\MeV)^2$ since with $q^2 = 2m_N E_R$,
where $E_R$ is the recoil energy of the nucleus, this is a typical value for
most SD detectors. Other momentum-suppressed operators can be handled the same
way.

\end{document}